\documentclass[lettersize]{IEEEtran}
\usepackage[utf8]{inputenc}

\usepackage{amsmath,amsfonts,amssymb,amsthm}
\usepackage{color}
\usepackage{tikz}
\usepackage[inline]{enumitem}
\usepackage{mdframed}
\usepackage{cleveref}
\usepackage{algorithm}
\usepackage[noend]{algpseudocode}
\usepackage{xspace}
\usepackage[caption=false]{subfig}
\usepackage{import}
\usepackage{wrapfig}

\newtheorem{theorem}{Theorem}
\newtheorem{theorem_repeat}{Theorem}

\definecolor{blueTileColor}{HTML}{afafff}
\newcommand{\blueTile}{{\color{blueTileColor}\blacksquare}}
\newcommand{\yellowTile}{{\color{yellow}\blacksquare}}
\definecolor{brownTileColor}{HTML}{f4a460} 
\newcommand{\brownTile}{{\color{brownTileColor}\blacksquare}}
\definecolor{redTileColor}{HTML}{ff8b8b} 
\newcommand{\redTile}{{\color{redTileColor}\blacksquare}}

\newenvironment{problembox}[1]{\begin{mdframed}[font=\small,backgroundcolor=black!10]\textbf{#1}}{\end{mdframed}}

\newcommand{\automaton}{{\mathcal{A}}}
\newcommand{\lang}{{\mathcal{L}}}

\newcommand{\problem}{\texttt{DFA-DIP}\xspace}

\begin{document}

\title{Learning Deterministic Finite Automata Decompositions from Examples and Demonstrations \thanks{This work was partially supported by NSF grants 1545126 (VeHICaL) and 1837132, by the DARPA contracts FA8750-18-C-0101 (Assured Autonomy) and FA8750-20-C-0156 (SDCPS), by Berkeley Deep Drive, by Toyota under the iCyPhy center, and by Toyota Research Institute.}}

\author{
  \IEEEauthorblockN{
    Niklas Lauffer\thanks{\textsuperscript{\IEEEauthorrefmark{1}}Equal contribution}\textsuperscript{\IEEEauthorrefmark{1}}\IEEEauthorrefmark{2},
    Beyazit Yalcinkaya\textsuperscript{\IEEEauthorrefmark{1}}\IEEEauthorrefmark{2},
    Marcell Vazquez-Chanlatte\IEEEauthorrefmark{2},
    Ameesh Shah\IEEEauthorrefmark{2} and
    Sanjit A. Seshia\IEEEauthorrefmark{2}
  }\\
  \IEEEauthorblockA{\IEEEauthorrefmark{2}University of California, Berkeley}
}

\maketitle


\begin{abstract}
The identification of a \emph{deterministic finite automaton} (DFA) from labeled examples is a well-studied problem in the literature; however, prior work focuses on the identification of monolithic DFAs.
Although monolithic DFAs provide accurate descriptions of systems’ behavior, they lack simplicity and interpretability; moreover, they fail to capture sub-tasks realized by the system and introduce inductive biases away from the inherent decomposition of the overall task.
In this paper, we present an algorithm for learning conjunctions of DFAs from labeled examples.
Our approach extends an existing SAT-based method to systematically enumerate Pareto-optimal candidate solutions.
We highlight the utility of our approach by integrating it with a state-of-the-art algorithm for learning DFAs from demonstrations.
Our experiments show that the algorithm learns sub-tasks realized by the labeled examples, and it is scalable in the domains of interest.
\end{abstract}

\section{Introduction}

Grammatical inference is a mature and well-studied field with many application domains ranging from various computer science fields, e.g., machine learning, to areas of natural sciences, e.g. computational biology~\cite{de2005bibliographical}. The identification of a minimum size \emph{deterministic finite automaton} (DFA) from labeled examples is one of the most well-investigated problems in this field. Furthermore, with the increase in computational power in recent years, the problem can be solved efficiently by various tools available in the literature (e.g.,~\cite{verwer2017flexfringe,zakirzyanov2019efficient}).

Existing work on DFA identification primarily focuses on the monolithic case, i.e., learning a single DFA from examples. Although such DFAs capture a language consistent with the examples, they may lack simplicity and interpretability. Furthermore, complex tasks often decompose into independent sub-tasks; hence, the system traces implicitly reflect this behavior. However, monolithic DFA identification fails to capture the natural decomposition of the system behavior, introducing an inductive bias away from the inherent decomposition of the overall task. In this paper, we present an algorithm for learning \emph{DFA decompositions} from examples by reducing the problem to graph coloring in SAT and a Pareto-optimal solution search over candidate solutions. A DFA decomposition is a set of DFAs such that \emph{intersection} of their language is the language of the system, which implicitly defines a conjunction of simpler specifications realized by the overall system. We present an application of our algorithm to a state-of-the-art method for learning task specifications from unlabeled demonstrations~\cite{marcell2021DISS} to showcase a domain of interest for DFA decompositions.



\textbf{Related Work.}
Existing work considers the problem of minimal DFA identification from labeled examples~\cite{de2005bibliographical}. It is shown that the DFA identification problem with a given upper bound on the number of states is an NP-complete problem~\cite{gold1978complexity}. Another work shows that this problem cannot be efficiently approximated~\cite{pitt1993minimum}. Fortunately, practical methods exist in the literature. A common approach is to apply the evidence driven state-merging algorithm~\cite{lang1998results,lang1999faster,bugalho2005inference}, which is a greedy algorithm that aims to find a good local optimum. Other works for learning DFAs use evolutionary computation~\cite{dupont1994regular,luke1999genetic}, later improved by multi-start random hill climbing~\cite{lucas2003learning}.

A different approach to the monolithic DFA identification is to leverage highly-optimized modern SAT solvers by encoding the problem in SAT~\cite{heule2010exact}. In follow up works, several symmetry breaking predicates are proposed for the SAT encoding to reduce the search space \cite{zakirzyanov2019efficient,ulyantsev2015bfs,ulyantsev2016symmetry,zakirzyanov2017finding}. However, to the best of our knowledge, no work considers directly learning DFA decompositions from examples and demonstrations.


This work also relates to the problem of decomposing a known automaton. Ashar et al.~\cite{ashar1992finite} explore computing cascade and general decomposition of finite state machines. The Krohn–Rhodes theorem \cite{rhodes2010applications} reduces a finite automaton into a cascade of irreducible automata.
Kupferman \& Mosheiff~\cite{kupferman2015prime} present various complexity results for DFA decomposability.

Finally, the problem of learning objectives from demonstrations of an expert dates back to the problem of Inverse Optimal Control~\cite{kalman1964linear} and, more recently in the artificial intelligence community, the problem of Inverse Reinforcement Learning (IRL)~\cite{ng2000irl}. The goal in IRL is to recover the unknown reward function that an expert agent is trying to maximize based on observations of that expert. Recently, several works have considered a version of the IRL problem in which the expert agent is trying to maximize the satisfaction of a Boolean task specification~\cite{kasenberg2017apprenticeship,chou2020explaining, marcell2021DISS}. However, no work considers learning \emph{decompositions} of specifications from demonstrations.

\section{Problem Formulation}

Let $\mathcal{D}$ denote the set of DFAs over some fixed alphabet $\Sigma$.
An $(m_1, \ldots, m_n)$-\emph{DFA decomposition} is a tuple of $n$ DFAs $(\automaton_1, \dots, \automaton_n) \in \mathcal{D}^n$ where $\automaton_i$ has $m_i$ states and $m_1 \leq m_2 \leq \dots \leq m_n$. We associate a partial order $\prec$ on DFA decompositions using the standard product order on the number of states. That is, $(\automaton_1', \dots, \automaton_n') \prec (\automaton_1, \dots, \automaton_n)$, if $m_i' \leq m_i$ for all $i \in [n]$ and $m_j' < m_j$ for some $j \in [n]$. In this case, we say $(\automaton_1', \dots, \automaton_n')$ \emph{dominates} $(\automaton_1, \dots, \automaton_n)$. A DFA decomposition $(\automaton_1, \ldots, \automaton_n)$ \emph{accepts} a string $w$ iff all $\automaton_i$ accept $w$. A string that is not accepted is \emph{rejected}. The \emph{language} of a decomposition, $\lang(\automaton_1, \ldots, \automaton_n)$, is the set of accepting strings, i.e., the intersection of all DFA languages.

We study the problem of finding a DFA decomposition from a set of positive and negative labeled examples such that the decomposition accepts the positive examples and rejects the negative examples. Next, we formally define \emph{the DFA decomposition identification problem} (\problem), and then present an overview of the proposed approach.
\begin{problembox}{The Deterministic Finite Automaton Decomposition Identification Problem (\problem).}
Given positive examples, $D_+$ and negative examples, $D_-$, and a natural number $n \in \mathbb{N}$, find a $(m_1, \ldots, m_n)$-DFA decomposition $(\automaton_1, \dots, \automaton_n)$ satisfying the following conditions. 
\begin{enumerate}[label=\textbf{(C\arabic*)}]
    \item\label{problem:language_condition} The decomposition is consistent with $(D_+, D_-)$:
    \[
    \begin{split}
        &D_+ \subseteq \lang(\automaton_1, \automaton_2, \dots, \automaton_n),\\
        &D_- \subseteq \Sigma^* \setminus \lang(\automaton_1, \automaton_2, \dots, \automaton_n).
    \end{split}
    \]
    \item\label{problem:numbers_of_states_condition} There does not exist a DFA decomposition that \emph{dominates} $(\automaton_1, \dots, \automaton_n)$ and satisfies \ref{problem:language_condition}.

\end{enumerate}
\end{problembox}

We refer to the set of DFA decompositions that solve an instance of \problem as the Pareto-optimal frontier of solutions. Note that for $n=1$, \problem reduces to monolithic DFA identification.
We propose finding the set of DFA decompositions that solve \problem by reduction to graph coloring in SAT and a breadth first search in solution space. Specifically, we extend the existing work on SAT-based monolithic DFA identification~\cite{heule2010exact,ulyantsev2016symmetry} to finding $n$ DFAs with $m_1, \dots, m_n$ states such that the intersection of their languages is consistent with the given examples.
On top of this SAT-based approach, we develop a search strategy over the numbers of states passed to the SAT solver as these values are not known a priori.



\section{Learning DFAs from Examples}

In this section, we present the proposed approach. We start with the SAT encoding of the DFA decomposition problem and continue with the Pareto frontier search in the solution space. We then showcase an example of learning conjunctions of DFAs from labeled examples. Finally, we present experimental results and evaluate the scalability of our method.

\subsection{Encoding \problem in SAT}

We extend the SAT encoding for monolithic DFA identification presented in~\cite{heule2010exact,ulyantsev2016symmetry}, which solves a graph coloring problem, to finding $n$ DFAs with $m_1, m_2, \dots, m_n$ states.
The extension relies on the observation that for conjunctions of DFAs, we need to enforce that a positive example must be accepted by \emph{all} DFAs, and a negative example must be rejected by \emph{at least} one of the DFAs.
Due to space limitations, we only present the modified clauses of the encoding, and invite reader to \Cref{sec:appendix} for further details.

The encoding works on an \emph{augmented prefix tree acceptor} (APTA), a tree-shaped automaton constructed from given examples, which has paths for each example leading to accepting or rejecting states based on the example's label; therefore, an APTA defines $D_+$ and $D_-$ which then constrains the accepting states, rejecting states, and the transition function of the unknown DFAs. For each DFA, $\automaton_i$, the encoding will associate the APTA states with one of the $m_i$ colors for DFA $\automaton_i$, subject to the constraints imposed by $D_+$ and $D_-$. APTA states with the same (DFA-indexed) color will be the same state in the corresponding DFA. We refer to states of an APTA as $V$, its accepting states as $V_{+}$, and its rejecting states as $V_{-}$. Given $n$ for the number of DFAs and $m_1, \dots, m_n$ for the number of states of DFAs, the SAT encoding uses three types of variables:
\begin{enumerate}
    \item \emph{color} variables $x^{k}_{v,i} \equiv 1$ ($k \in [n]$; $v \in V$; $i \in [m_k]$) iff APTA state $v$ has color $i$ in DFA $k$,
    \item \emph{parent relation} variables $y^{k}_{l,i,j} \equiv 1$ ($k \in [n]$; $l \in \Sigma$, where $\Sigma$ is the alphabet; $i, j \in [m_k]$) iff DFA $k$ transitions with symbol $l$ from state $i$ to state $j$, and
    \item \emph{accepting color} variables $z^{k}_{i} \equiv 1$ ($k \in [n]$; $i \in [m_k]$) iff state $i$ of DFA $k$ is an accepting state.
\end{enumerate}
The encoding for the monolithic DFA identification also uses the same variable types; however, in our encoding, we also index variables over $n$ DFAs instead of a single DFA. With this extension, one can trivially instantiate the encoding presented in \cite{heule2010exact,ulyantsev2016symmetry}. Below, we list the new rules we define for our problem. For the complete list of rules, see~\Cref{sec:appendix}.
\begin{enumerate}[label=\textbf{(R\arabic*)},leftmargin=23pt]
    \item\label{rule1} A negative example must be rejected by \emph{at least} one DFA:
    \[
        \bigwedge_{v \in V_{-}} \bigvee_{k \in [n]} \bigwedge_{i \in [m_k]} x^{k}_{v, i} \implies \neg z^{k}_{i}.
    \]
    \item\label{rule2} Accepting and rejecting states of APTA cannot be merged:
    \[
        \bigwedge_{v_{-} \in V_{-}} \bigwedge_{v_{+} \in V_{+}} \bigwedge_{k \in [n]} \bigwedge_{i \in [m_k]} (x^{k}_{v_{-}, i} \wedge \neg z^{k}_{i}) \implies \neg x^{k}_{v_{+}, i}.
    \]
\end{enumerate}
In the encoding of \cite{heule2010exact,ulyantsev2016symmetry}, we replace the rule stating that the resulting DFA must reject all negative examples with \ref{rule1}, and \ref{rule2} is used instead of the original rule stating that accepting and rejecting states of APTA cannot be merged. Notice that since a rejecting state of APTA is not necessarily a rejecting state of a DFA $k$, we need to use the new rule~\ref{rule2}.

\begin{theorem}\label{thm:sat}
Given labeled examples with $n$ and $m_1, \dots, m_n$, a solution to our SAT encoding is a solution to \problem.
\end{theorem}

See \Cref{appendix:proof_main} for the proof of \Cref{thm:sat}.

\begin{figure*}[t]
    \centering
  \subfloat[Experiment results answering \ref{q1}, where we vary number of DFAs.\label{fig:vary_dfas}]{%
       \includegraphics[width=0.49\linewidth]{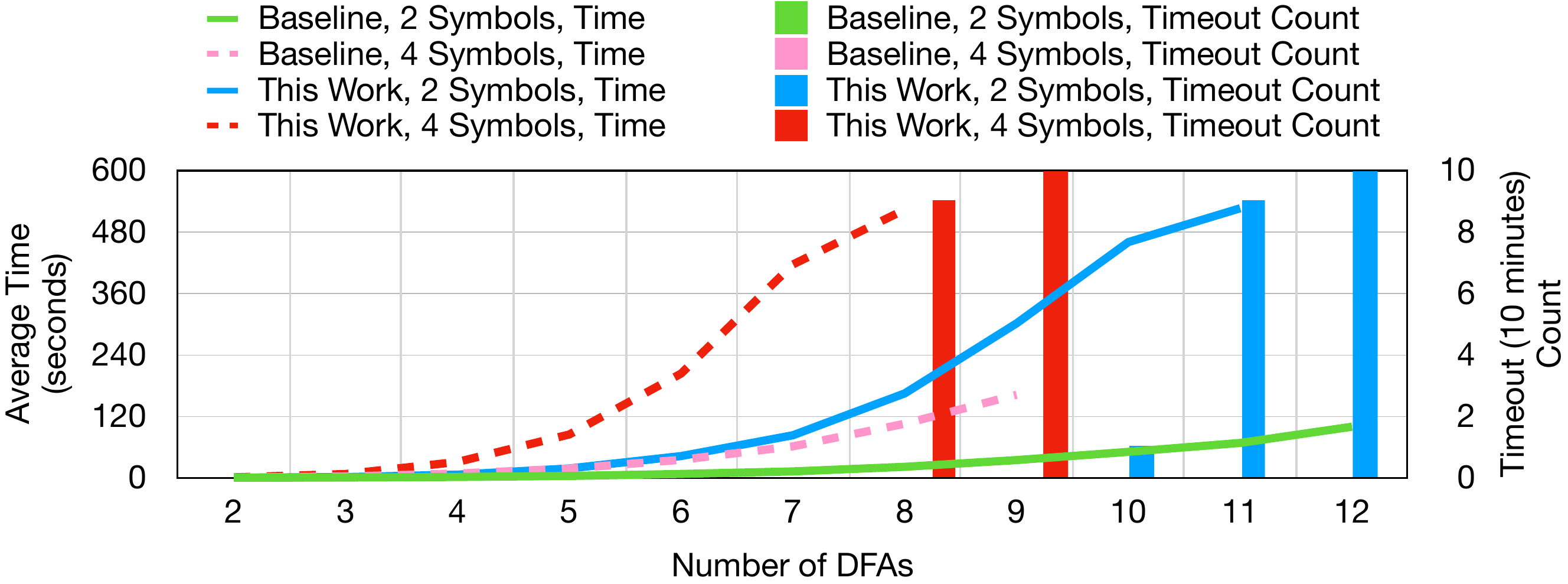}}
    \hfill
  \subfloat[Experiment results answering \ref{q2}, where we vary number of examples.\label{fig:vary_examples}]{%
       \includegraphics[width=0.49\linewidth]{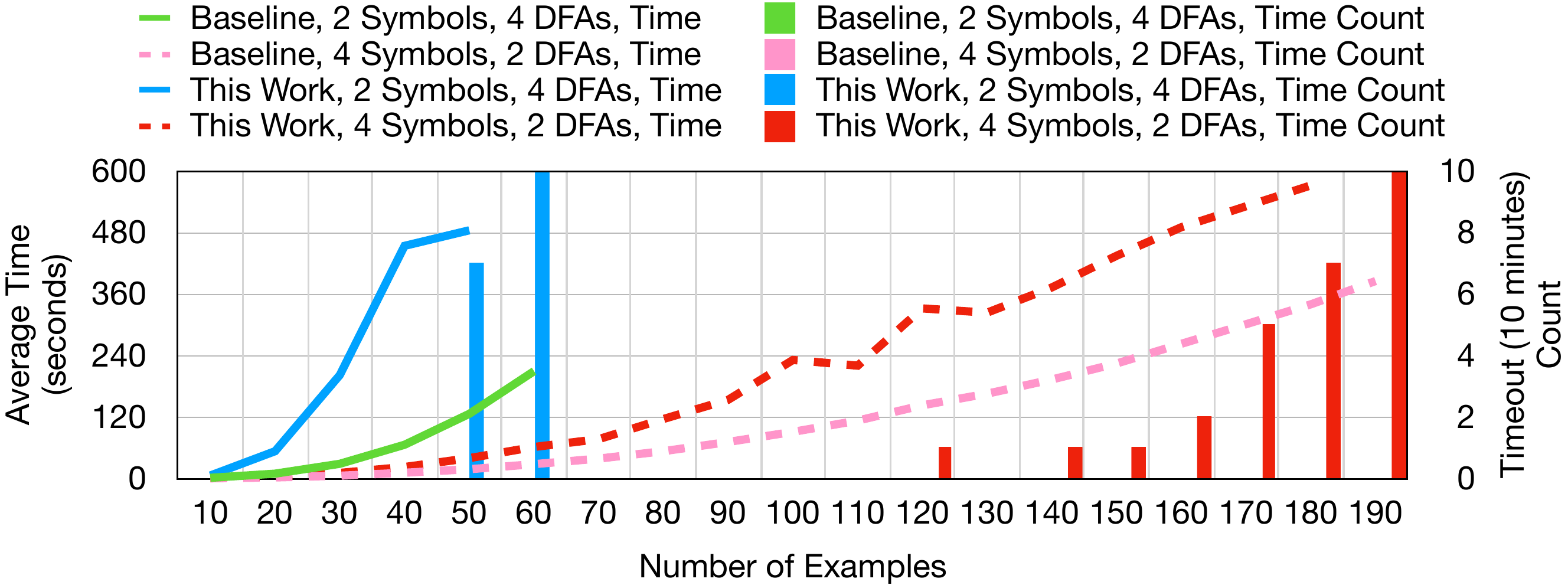}}
  \caption{Experiment results evaluating the scalability of our algorithm w.r.t. (a) number of DFAs implied by the examples and (b) number of labeled examples.}
  \label{fig:exps}
\end{figure*}



\subsection{Pareto Frontier Search}

\problem requires finding a conjunction of $n$ DFAs that identify a language. There may exist multiple DFA decompositions that solve the problem with varying number of states $m_1, m_2, \dots, m_n$. With only a single DFA, the notion of minimal size is well-captured by the number of states. However, with multiple DFAs in the decomposition, the notion of a minimal solution is less clear. For example, there may exist a decomposition of two size three DFAs, and a separate decomposition of a size two and a size four DFA, both identifying the given set of labeled examples. Neither solution is strictly smaller than the other, and either solution might be preferred in different scenarios. Therefore, the set of solutions to \problem form a Pareto-optimal frontier in solution space.

Our proposed Pareto frontier enumeration algorithm is a breadth first search (BFS) over DFA decomposition size tuples that skips tuples that are dominated by an existing solution. 
This BFS is over a directed acyclic graph $G = (V,E)$ formed in the following way. There is a vertex in the graph for every ordered tuple of states sizes. There is  an edge from $(m_1, m_2, \dots, m_n)$ to $(m'_1, m'_2, \dots, m'_n)$ if there exists some $j \in [n]$ such that:
\begin{equation*}
    m'_i = 
    \begin{cases}
        m_i + 1 & \text{if $i = j$;} \\
        m_i & \text{otherwise.}
    \end{cases}
\end{equation*}
A size tuple $(m_1, \dots, m_n)$ is a sink, i.e., the search does not continue past this vertex, if there exists a $(m_1, \dots, m_n)$-decomposition that solves \problem or the size tuple is dominated by a previously traversed solution. In the prior case, the associated DFA decomposition is also returned as a solution on the Pareto-optimal frontier.
The BFS starts from $m_1 = m_2 = \dots = m_n = 1$, and performs the search as explained.
See \Cref{appendix:pareto_search} for the details of the algorithm.

\begin{theorem}\label{thm:pareto}
The described BFS is sound and complete; it outputs the full Pareto-optimal frontier of solutions without returning any dominated solutions.
\end{theorem}

See \Cref{appendix:proof_pareto} for the proof of \Cref{thm:pareto}.

\subsection{Example: Learning Partially-Ordered Tasks}\label{subsec:toy_example}

We continue with a toy example showcasing the capabilities of the proposed approach. Later, we use the same class of decompositions to evaluate the scalability of our algorithm.
\setlength{\columnsep}{10pt}
\begin{wrapfigure}[13]{l}{0.25\textwidth}
\centering
\vspace{-20pt}
\subfloat[Learned DFA recognizing the ordering between $\yellowTile$ and $\redTile$.\label{fig:toy_example_dfa1}]{%
       \includegraphics[width=\linewidth]{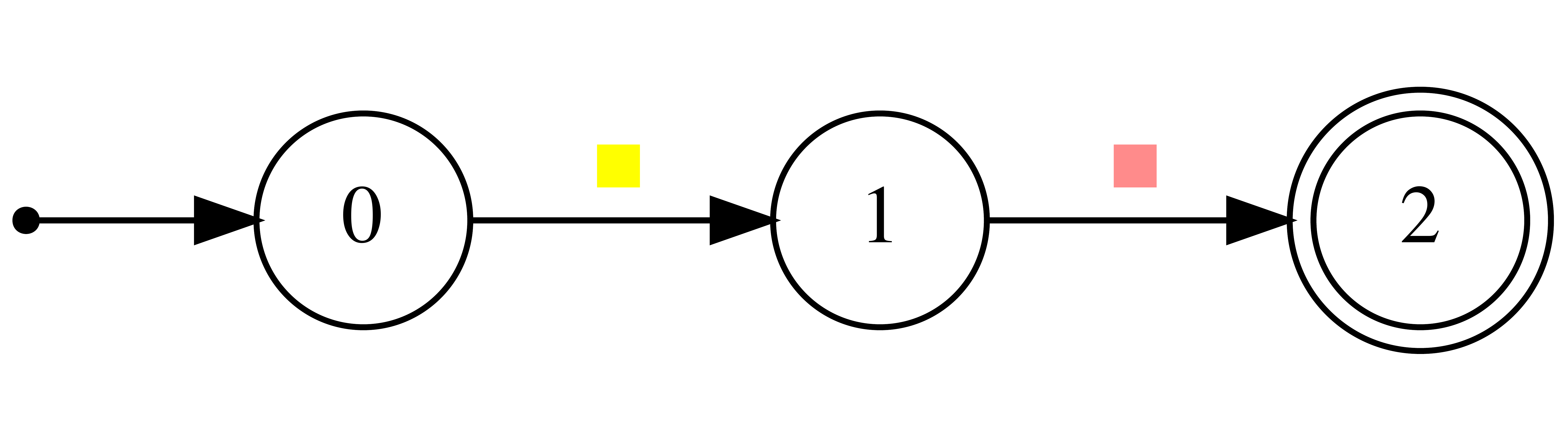}}
    \vfill
  \subfloat[Learned DFA recognizing the ordering between $\blueTile$ and $\brownTile$.\label{fig:toy_example_dfa2}]{%
       \includegraphics[width=\linewidth]{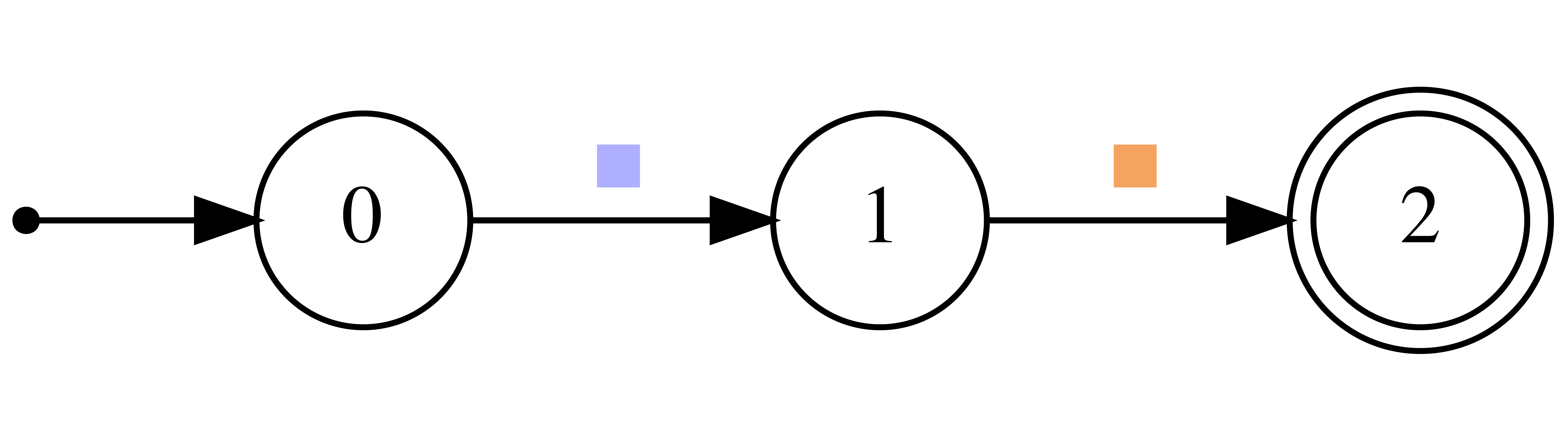}}
\caption{Learned DFA decomposition.}
\label{fig:toy_example} 
\end{wrapfigure}
Inspired from the multi-task reinforcement learning literature~\cite{vaezipoor2021ltl2action}, our example focuses on partially-ordered temporal tasks executed in parallel. 
Specifically, consider a case where an agent is performing two ordering tasks in parallel:
\begin{enumerate*}[label=(\roman*)]
    \item observe $\yellowTile$ before $\redTile$, and
    \item observe $\blueTile$ before $\brownTile$.
\end{enumerate*}
A positive example of such behavior is simply any sequence of observations ensuring both of the given orderings, e.g. $\yellowTile\redTile\blueTile\brownTile$, and a negative example is any sequence that fails to satisfy both orderings, e.g. $\yellowTile\redTile\brownTile\blueTile$. We generate such positive and negative examples and feed them to our algorithm. \Cref{fig:toy_example} presents the learned DFAs recognizing ordering sub-tasks of the example. The intersection of their languages is consistent with the given observations, and their conjunction is the overall task realized by the system generating the traces. The monolithic DFA recognizing the same language has nine states, and is more complicated, see \Cref{fig:monolithic_dfa} in \Cref{appendix:figs}.




\subsection{Experimental Evaluation}\label{subsec:dfa_eval}

We evaluate the scalability of our algorithm through experiments with changing sizes of partially-ordered tasks introduced in \Cref{subsec:toy_example}. In our evaluation, we aim to answer two questions:
\begin{enumerate*}[label=\textbf{(Q\arabic*)}]
    \item\label{q1} ``How does solving time scale with the number of ordering tasks?'', and
    \item\label{q2} ``How does solving time scale with the number of labeled examples?''.
\end{enumerate*}
We implement our algorithm in Python with PySAT~\cite{imms-sat18}, and we use Glucose4~\cite{een2003extensible} as the SAT solver. Our baseline is an implementation of the monolithic DFA identification encoding from~\cite{heule2010exact,ulyantsev2016symmetry} with the same software as our implementation. Experiments are performed on a Quad-Core Intel i7 processor clocked at 2.3 GHz and a 32 GB main memory.

To evaluate the scalability, we randomly generate positive and negative examples with varying problem sizes. For \ref{q1}, we generate 10 (half of which is positive and half of which is negative) partially-ordered task examples with (i) 2 symbols, and (ii) 4 symbols, and we vary the number of DFAs from 2 to 12. For \ref{q2}, we generate 10 to 20  partially-ordered task examples  with (i) 2 symbols and 4 DFAs, and (ii) 4 symbols and 2 DFAs. Half of these examples are positive and the other half is negative. Since the examples are generated randomly, we run the experiments for 10 different random seeds and report the average. We set the timeout limit to 10 minutes, and stop when our algorithm timeouts for all random seeds.

\begin{figure*}[t] \label{fig:diss_exp}
    \centering
    \parbox{\textwidth}{
        \parbox{.28\textwidth}{%
          \subfloat[A stochastic grid world environment with expert demonstrations of an agent trying to accomplish a task.]{\label{fig:gridworld}
            \centering \scalebox{.9}{
            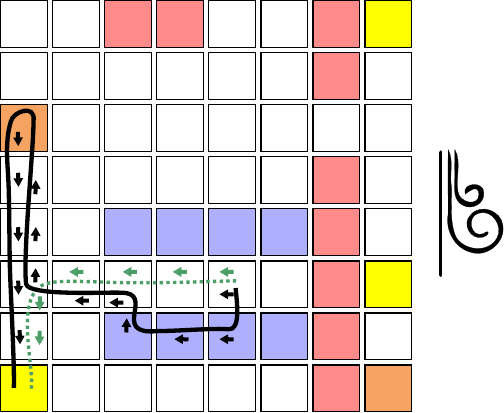}
            }
        }
        \parbox{.22\textwidth}{%
          \subfloat[Labeled examples conjectured by DISS.]{\begin{tabular}{l|l}
            Positive & Negative \\
            \hline
            $\yellowTile$  & $\blueTile$ \\[-4pt]
            $\yellowTile \yellowTile$ & $\redTile$ \\[-4pt]
            $\yellowTile \blueTile$ & $\blueTile \redTile$ \\[-4pt]
            $\blueTile \brownTile \yellowTile$ & $\blueTile \brownTile$ \\[-4pt]
            & $\blueTile \yellowTile$ \\[-4pt]
            & $\redTile \brownTile$ \\[-4pt]
            & $\redTile \blueTile \redTile$ \\[-4pt]
            & $\blueTile \redTile \yellowTile$ \\[-4pt]
            & $\blueTile \redTile \brownTile \yellowTile$ \\[-4pt]
            & $\redTile \yellowTile \brownTile \redTile$ \\[-4pt]
            & $\blueTile \redTile \blueTile \redTile \yellowTile$ \\[-4pt]
            & $\redTile \yellowTile \redTile \yellowTile \redTile$ \\[-4pt]
            & $\redTile \blueTile \redTile \blueTile \redTile \yellowTile$
          \end{tabular} \label{fig:diss_exp:examples}}
        }
        \parbox{.21\textwidth}{%
            \subfloat[Go to $\yellowTile$.\label{fig:dfa_diss1}]{\includegraphics[width=0.2\textwidth]{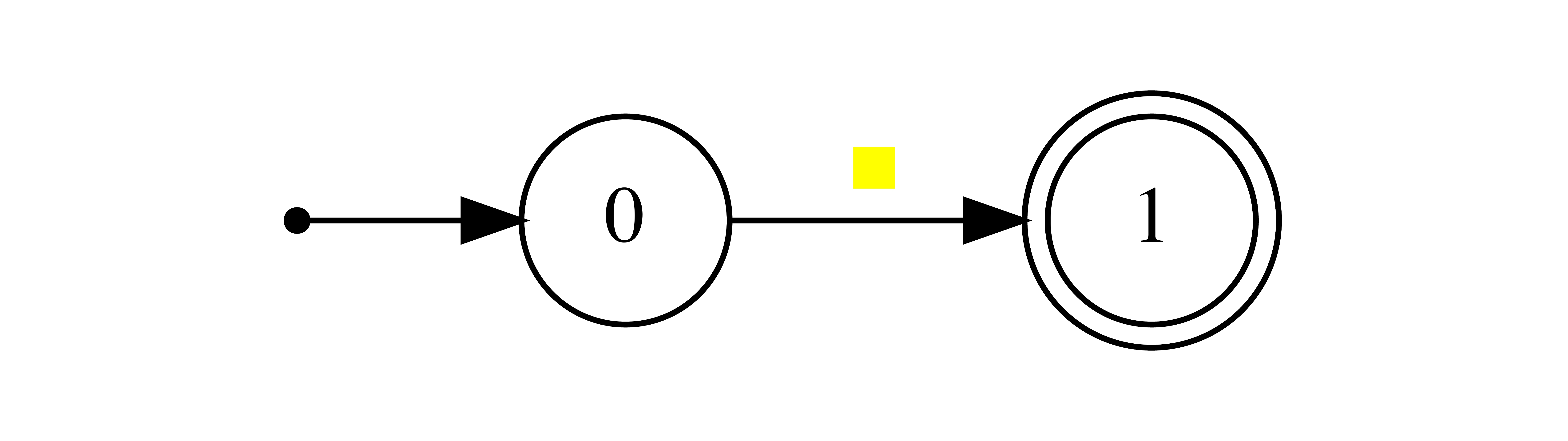}}
            \vspace{1pt}
            \subfloat[Avoid $\redTile$.\label{fig:dfa_diss2}]{\includegraphics[width=0.2\textwidth]{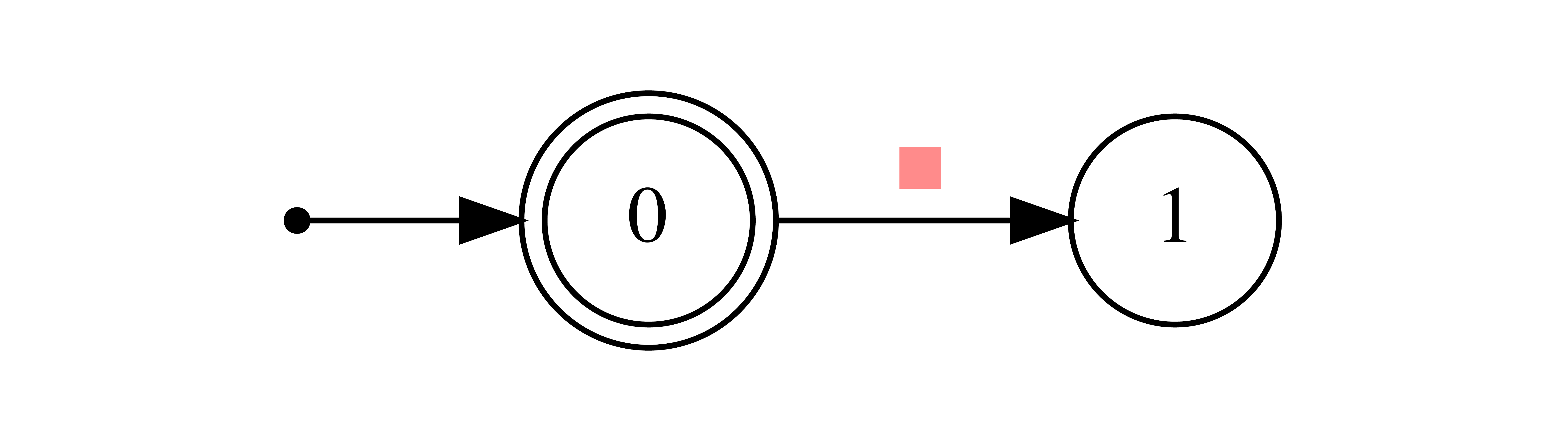}}
            \vspace{1pt}
            \subfloat[After $\blueTile$, go to $\brownTile$ before $\yellowTile$.\label{fig:dfa_diss3}]{\includegraphics[width=0.2\textwidth]{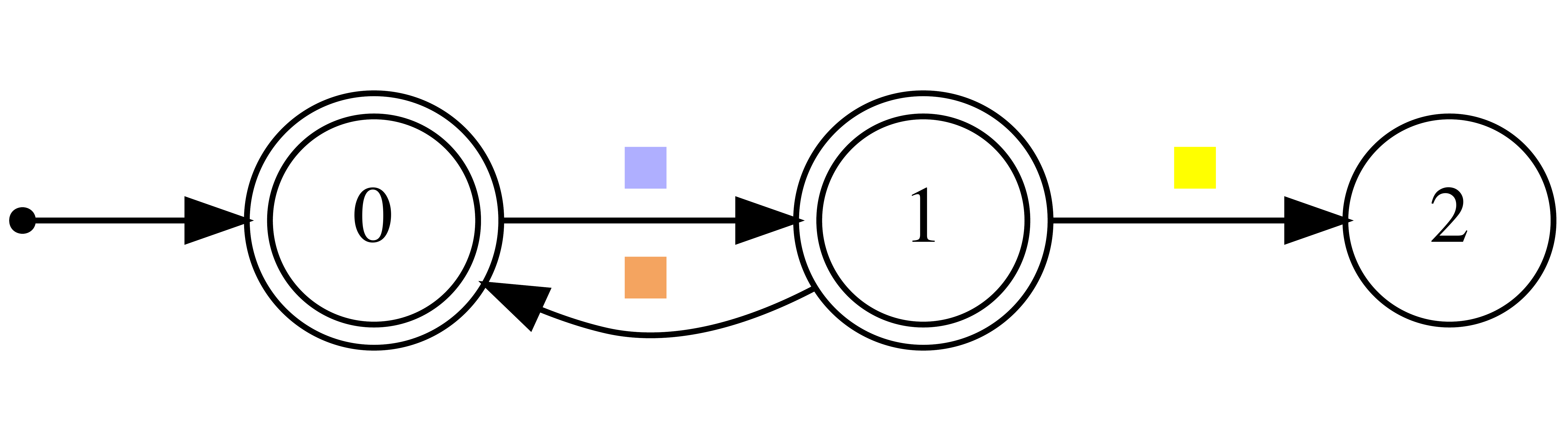}}
        }
        \parbox{.27\textwidth}{%
          \subfloat[Monolithic DFA for the example presented in \Cref{subsec:toy_example}.]{\label{fig:diss}
            \includegraphics[width=0.25\textwidth]{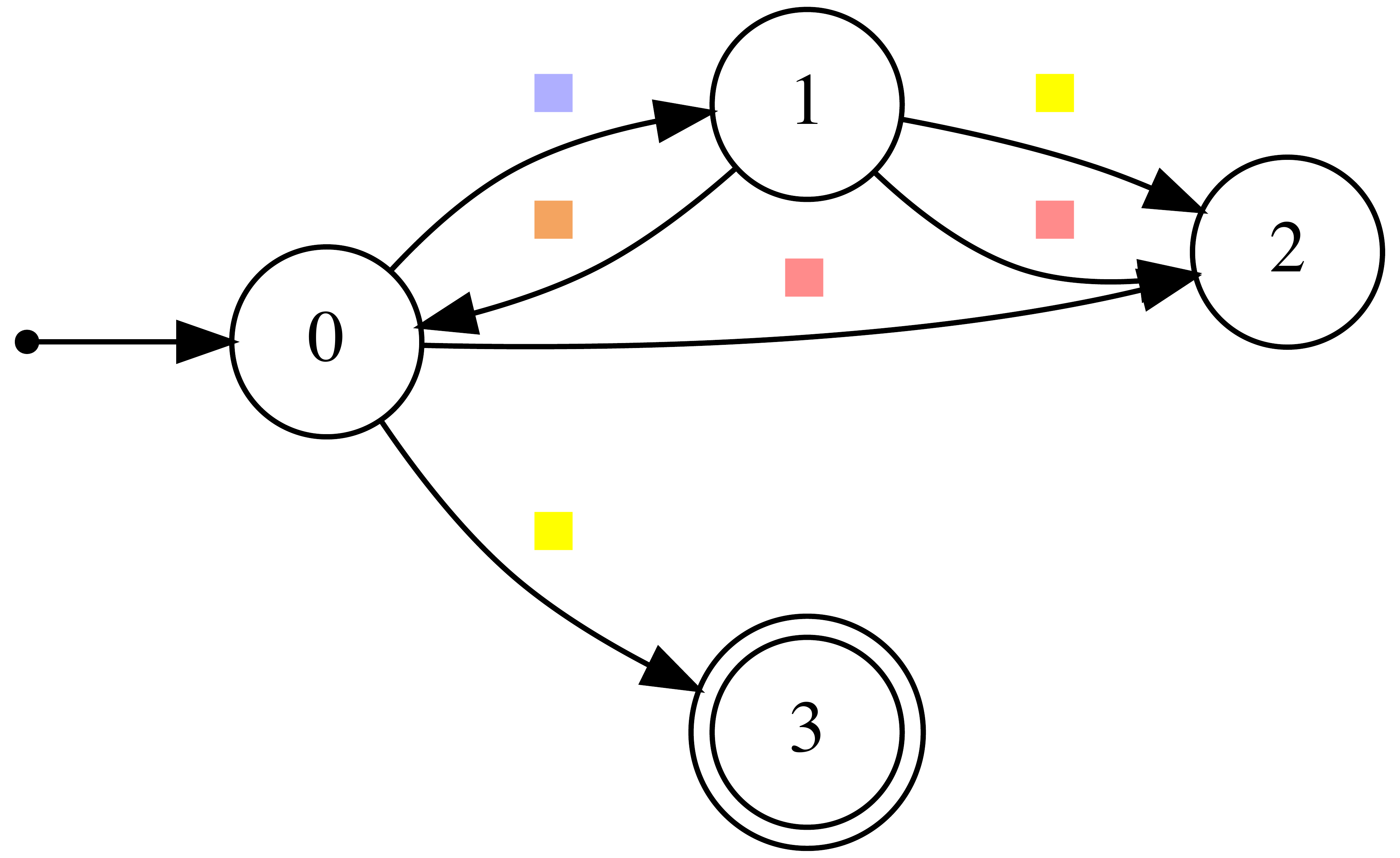}
            }
        }
      }
        
     \caption{\Cref{fig:gridworld} shows the stochastic grid world environment. \Cref{fig:diss_exp:examples} shows the positive and negative examples of the expert's behavior conjectured by DISS and \Cref{fig:dfa_diss1,fig:dfa_diss2,fig:dfa_diss3} showcases the associated DFA decomposition identified by our algorithm. \Cref{fig:diss} shows the monolithic DFA learned in~\cite{marcell2021DISS}.}
\end{figure*}

\Cref{fig:vary_dfas} presents the experiment results answering \ref{q1}, where we vary the number of DFAs implied by the given examples. For partially-ordered tasks with 2 symbols, green solid line is the (monolithic DFA) baseline and the blue solid is our algorithm. Similarly, for partially-ordered tasks with 4 symbols, pink dashed line is the baseline and the red dashed line is our algorithm. \Cref{fig:vary_examples} presents the experiment results answering \ref{q2}, where we vary the number of examples. For partially-ordered tasks with 2 symbols and 4 DFAs, green solid line is the baseline and the blue solid is our algorithm; for partially-ordered tasks with 4 symbols and 2 DFAs, pink dashed line is the baseline and the red dashed line is our algorithm. As expected, the baseline scales better than our algorithm as we also search for the Pareto frontier and solve an inherently harder problem. Notice that given 10 examples, our algorithm is able to scale up to 11 DFAs for tasks with 2 symbols, and 8 DFAs for tasks with 4 symbols; for 2 symbols and 4 DFAs, it is able to scale up to 60 examples, and for 4 symbols and 2 DFAs, it is able to scale up to 190 examples. As we demonstrate in the next section, these limits for scalability are practically useful in certain domains.


\section{Learning DFAs from Demonstrations}

Next, we show how our algorithm can be incorporated into Demonstration Informed Specification Search (DISS) - a framework for learning languages from expert demonstrations~\cite{marcell2021DISS}. For our purposes a \emph{demonstration} is an unlabeled path through a workspace that maps to a string and is biased towards being accepting by some unknown language.
For example, we ran our implementation of DISS using demonstrations produced by an expert attempting
to accomplish a task in a stochastic grid world environment, the same example used in \cite{marcell2021DISS} and shown in~\Cref{fig:gridworld}. At each step, the agent can move in any of the four cardinal directions, but because of wind blowing from the north to the south, with some probability, the agent will transition to the space south of it in spite of its chosen action.
Two demonstrations of the task ``Reach $\yellowTile$ while avoiding $\redTile$. If it ever touches $\blueTile$, it must then touch $\brownTile$ before reaching $\yellowTile$.'' are shown in~\Cref{fig:gridworld}. 

In order to efficiently search for tasks, DISS reduces the learning from demonstrations problem into a series of identification problems to be solved by a black-box identification algorithm. The goal of DISS is to find a task that minimizes the joint description length, called the energy, of the task and the demonstrations assuming the agent were performing said task. The energy is measured in bits to encode an object.

Below, we reproduce the results from~\cite{marcell2021DISS}, but using our algorithm as the task identifier rather than the monolithic DFA identifier provided.
The use of DFA decompositions biases DISS to conjecture concepts that are \emph{simpler} to express in terms of a DFA decomposition. 
To define the description length of DFA decompositions, we adapt the DFA encoding used in~\cite{marcell2021DISS} by expressing a decomposition as the concatenation of the encodings of the individual DFAs. To remove unnecessary redundancy two optimizations were performed. First common headers, e.g. indicating the alphabet size, were combined. Second, as the DFAs in a decomposition are ordered by size, we expressed changes in size rather than absolute size, see \Cref{appendix:encode_sizes} for details.



\subsection{Experimental Evaluation}\label{subsec:diss_eval}


In~\Cref{fig:dfa_diss1,fig:dfa_diss2,fig:dfa_diss3} we present the learned DFA decomposition along with the corresponding~\Cref{fig:diss_exp:examples} labeled examples conjectured by DISS to explain the expert behavior. Importantly, this decomposition exactly captures the demonstrated task. We note that this is in contrast to the DFA learned in~\cite{marcell2021DISS}, shown in~\Cref{fig:diss}, which allows visiting $\redTile$ after visiting $\yellowTile$. Further, we remark that the time required to learn the monolithic and decomposed DFAs was comparable. In particular, the number of labeled examples was less than 60 and as with the monolithic baseline, most of the time is not spent in task identification, but instead conjecturing the labeled examples. As we saw with in~\Cref{subsec:dfa_eval}, this number of examples is easily handled by our SAT-based identification algorithm. Finally, the number of labeled examples that needed to be conjectured to find low energy tasks was similar for both implementations (see~\Cref{fig:diss_exp:plot,fig:diss_exp:inc_plot} in~\Cref{appendix:figs}). Thus, our variant of DISS performed similar to the monolithic variant, while finding DFAs that exactly represented the task.


\section{Conclusion}
To the best of our knowledge, this work presents the first approach for solving \problem. Our algorithm works by reducing the problem to a Pareto-optimal search of the space the number of states in a DFA decomposition with a SAT call in the inner loop.
The SAT-based encoding is based on an efficient reduction to graph coloring.
We demonstrated the scalability of our algorithm on a class of problems inspired by the multi-task reinforcement learning literature and show that the additional computational cost for identifying DFA decompositions over monolithic DFAs is not prohibitive. Finally, we showed how identifying DFA decompositions can provide a useful inductive bias while learning from demonstrations.

\bibliographystyle{IEEEtran}
\bibliography{references}

\appendix\label{sec:appendix}

\subsection{Complete SAT Encoding of \problem}
Below, we list the complete SAT encoding of the \problem. Observe that the encoding extends the SAT encoding for monolithic DFA identification presented in~\cite{heule2010exact,ulyantsev2016symmetry}. We refer to the root node (i.e., the initial state) of an APTA as $v_r$, for $v \in V \setminus \{v_r\}$, $l(v)$ denotes the symbol on the incoming transition of $v$, and $p(v)$ is the parent node (i.e., the previous state) of $v$, as APTA is a tree-like automaton, $p(v)$ is unique.

\begin{enumerate}
    \item A positive example must be accepted by \emph{all} DFAs:
    \[
        \bigwedge_{v \in V_{+}} \bigwedge_{k \in [n]} \bigwedge_{i \in [m_k]} x^{k}_{v, i} \implies z^{k}_{i}.
    \]
    \item A negative example must be rejected by \emph{at least} one DFA:
    \[
        \bigwedge_{v \in V_{-}} \bigvee_{k \in [n]} \bigwedge_{i \in [m_k]} x^{k}_{v, i} \implies \neg z^{k}_{i}.
    \]
    \item Each state of APTA has at least one color for each DFA:
    \[
        \bigwedge_{v \in V} \bigwedge_{k \in [n]} \bigwedge_{i \in [m_k]} x^{k}_{v, i}.
    \]
    \item A transition of a DFA is set when a state and its parent are both colored:
    \[
        \bigwedge_{v \in V \setminus \{v_r\}} \bigwedge_{k \in [n]} \bigwedge_{i,j \in [m_k]} (x^{k}_{p(v), i} \wedge x^{k}_{v, j}) \implies y^{k}_{l(v), i, j}.
    \]
    \item A transition of a DFA targets at most one state:
    \[
        \bigwedge_{l \in \Sigma} \bigwedge_{k \in [n]} \bigwedge_{\substack{i, j, t \in [m_k]\\j < t}} y^{k}_{l, i, j} \implies \neg y^{k}_{l, i, t}.
    \]
    \item Each state of APTA has at most one color for each DFA:
    \[
        \bigwedge_{v \in V} \bigwedge_{k \in [n]} \bigwedge_{i, j \in [m_k]} \neg x^{k}_{v, i} \vee \neg x^{k}_{v, j}.
    \]
    \item A transition of a DFA targets at least one state:
    \[
        \bigwedge_{l \in \Sigma} \bigwedge_{k \in [n]} \bigwedge_{i,j \in [m_k]} y^{k}_{l, i, j}.
    \]
    \item For each DFA, a node color is set when the color of the parent node and the transition between them are set:
    \[
        \bigwedge_{v \in V \setminus \{v_r\}} \bigwedge_{k \in [n]} \bigwedge_{i, j \in [m_k]} (x^{k}_{p(v), i} \wedge y^{k}_{l(v), i, j}) \implies x^{k}_{v, j}.
    \]
    \item Accepting-rejecting nodes of APTA cannot be merged:
    \[
        \bigwedge_{v_{-} \in V_{-}} \bigwedge_{v_{+} \in V_{+}} \bigwedge_{k \in [n]} \bigwedge_{i \in [m_k]} (x^{k}_{v_{-}, i} \wedge \neg z^{k}_{i}) \implies \neg x^{k}_{v_{+}, i}.
    \]
\end{enumerate}

The next set of constraints encode the symmetry breaking clauses intruduced in~\cite{ulyantsev2016symmetry} to avoid consideration of isomorphich DFAs. The main idea of the symmetry breaking clauses is to enforce individual DFA states to be enumerated in a depth-first search (DFS) order. See~\cite{ulyantsev2016symmetry} for more details. The symmetry breaking clauses make use of new auxilliary variables $p^k_{j,i}$ and $t^k_{i,j}$ for $k \in [n], i,j \in [m_k]$ and $m_{l,i,j}$ for $l \in \Sigma, i,j \in [m_k]$. Let $\Sigma = \{l_1, \dots, l_L\}$.
\begin{enumerate}
    \item Each state must have a smaller parent in the DFS order:
    \[
      \bigwedge_{k \in [n]} \bigwedge_{i \in [2,m_k]} ( p^{k}_{i, 1} \lor \dots \lor p^{k}_{i,i-1} ).
    \]
  \item Define $p^k_{j,i}$ in terms of auxilliary variable $t^k_{i,j}$:
    \[
      \bigwedge_{k \in [n]} \bigwedge_{\substack{i,j \in [m_k] \\ i < j}} (p^k_{j,i} \iff t^k_{i,j} \land t^k_{i+1,j} \land \dots \land t^k_{j-1,j} ) 
    \]
  \item Define $t^k_{i,j}$ in terms of $y_{l,j,j}$:
    \[
      \bigwedge_{k \in [n]} \bigwedge_{\substack{i,j \in [m_k] \\ i < j}} (t^k_{i,j} \iff y^k_{l_1,i,j} \lor \dots \lor y^k_{l_L,i,j})
    \]
  \item The parent relationship follows the DFS order
    \[
      \bigwedge_{k \in [n]} \bigwedge_{\substack{i,j,p,q \in [m_k] \\ i < p < j < q}} (p^k_{j,i} \implies \lnot t^k_{p,q})
    \]
  \item Define $m^k_{l,i,j}$ in terms of $y^k_{l,i,j}$:
    \[
      \bigwedge_{k \in [n]}  \bigwedge_{\substack{i,j \in [m_k] \\ i < j}} \bigwedge_{l_r \in \Sigma} (m^k_{l_r,i,j}  \iff y^k_{l_r,i,j} 
      \dots \land y^k_{l_1,i,j})
    \]
  \item Enforce DFAs to be DFS-enumerated in the order of symbols on transitions:
    \[
      \bigwedge_{k \in [n]} \bigwedge_{\substack{i,j,q \in [m_k] \\ i < j < q}} \bigwedge_{\substack{l_r, l_s \in \Sigma \\ r < s}} (p^k_{j,i} \land p^k_{q,i} \land m^k_{l_r,i,}j \implies \lnot m^k_{l_s,i,k})
    \]
\end{enumerate}


\subsection{Proof of \Cref{thm:sat}} \label{appendix:proof_main}

\begin{theorem_repeat}
Given labeled examples with $n$ and $m_1, \dots, m_n$, a solution to our SAT encoding is a solution to \problem.
\end{theorem_repeat}



\begin{IEEEproof}
  We assume that the SAT-based reduction to graph coloring for monolithic DFA identification given in \cite{heule2010exact} is correct.
  Constraint \ref{rule1} and \ref{rule2} replace similar constraint in the monolithic encoding given in \cite{heule2010exact}:
\begin{enumerate}[label=\textbf{(R\arabic*')},leftmargin=26pt]
    \item\label{rule1_prime} a negative example must be rejected by the DFA:
    \[
        \bigwedge_{v \in V_{-}} \bigwedge_{i \in [m_k]} x_{v, i} \implies \neg z_{i}, \text{ and}
    \]
    \item\label{rule2_prime} accepting and rejecting states of the APTA cannot be merged:
    \[
        \bigwedge_{v_{-} \in V_{-}} \bigwedge_{v_{+} \in V_{+}} \bigwedge_{i \in [m_k]} x_{v_{-}, i} \implies \neg x_{v_{+}, i}.
    \]
\end{enumerate}
In the monolithic DFA case, there is only a single DFA so for ease of notation, we drop the index $k$.
First notice that constraints \ref{rule1_prime} and \ref{rule2_prime} have no bearing on whether the DFA accepts each positive example. Therefore, our encoding automatically requires that each DFA in the DFA decomposition accepts all of the positive examples and is not constrained to unecessarily accept any unspecified examples.

Constraint \ref{rule1_prime} ensures that the resulting monolithic DFA rejects every negative example by making the color of the node in the APTA associated with the negative example rejecting.
Constraint \ref{rule1} replaces this and ensures that at least one of the DFAs in the DFA decomposition rejects a negative example by making the color of the node in the APTA associated with the negative example rejecting in at least one of the $n$ DFAs in the decomposition. Thus, the language intersection of the resulting decomposition correctly rejects negative examples.

Constraint \ref{rule2_prime} ensures that all pairs of rejecting and accepting nodes of the APTA cannot be assigned the same color (i.e., merged) in the resulting DFAs. Constraint \ref{rule2}, which replaces \ref{rule2_prime}, ensures that for each DFA in the decomposition, the pair $(x_{v_{-},i}^k, x_{v_{+},i}^k)$ of accepting and rejecting nodes of the APTA cannot be assigned the same color only if DFA $k$ is rejecting the negative example associated with $x_{v_{-},i}^k$ (which is handled by constraint \ref{rule1}).
This allows all but one DFA in the DFA decomposition to accept negative examples. Therefore, no DFA in the decomposition is constrained to unnecessarily reject a negative example if some other DFA in the DFA decomposition already does so. Therefore, the language intersection of the DFAs in the DFA decomposition is not constrained to reject any unspecified examples.

So, if there exists a DFA decomposition with the specified number of states such that all DFAs accept the positive examples and at least one DFA in the decomposition rejects each rejecting example, our encoding will find it.
\end{IEEEproof}

\subsection{Details of the Pareto Frontier Search Algorithm} \label{appendix:pareto_search}

\Cref{alg:pareto_frontier} presents the details of the BFS performed in the solution space for finding the Pareto-optimal frontier.

\begin{algorithm}[h]
  \caption{Pareto frontier enumeration algorithm.
    \label{alg:pareto_frontier}}
  \begin{algorithmic}[1]
    \Require{Positive $D_{+}$ and negative $D_{-}$ labeled examples and positive integer $n$.}
    \State $Q^{\star} \gets \emptyset$ \Comment{Maintains the Pareto frontier}
    \State $S \gets \{(2,\dots,2)\}$ \Comment{Initialize the queue}
    \While{$S \neq \emptyset$}
        \State $m \gets Q.dequeue()$
        \If{$\nexists \hat{m} \in P^{\star} \text{ s.t. } \hat{m} \prec m$}
            \State $SAT, \mathcal{A} \gets \textproc{Solve}(n, m, D_{+}, D_{-})$
            \If{$SAT$}
                \State $P^{\star} = P^{\star} \cup \mathcal{A}$ \Comment{Add to the Pareto frontier}
                \Else
                \For{$k=1,\dots,n$}
                    \State $m' \gets m$
                    \State $m'_k \gets m'_k + 1$
                    \If{$\text{ordered}(m')$}
                        \State $Q.enqueue(m')$
                    \EndIf
                \EndFor
            \EndIf
        \EndIf
    \EndWhile
    \State \Return{$P^{\star}$}
  \end{algorithmic}
\end{algorithm}

\subsection{Proof of \Cref{thm:pareto}} \label{appendix:proof_pareto}

\begin{theorem_repeat}
The described BFS is sound and complete; it outputs the all Pareto-optimal frontier of solutions without returning any dominated solutions.
\end{theorem_repeat}



\begin{IEEEproof}
The described BFS enumerates the number of states of DFA decomposition in product order. Therefore, before reaching a vertex with number of states $(m_1, m_2, \dots, m_n)$, it explores all number of states of DFA decompositions that dominate the DFA decomposition with $(m_1, m_2, \dots, m_n)$ states. If any of these number of states admit a solution to \problem, then the DFA decomposition associated with number of states $(m_1, m_2, \dots, m_n)$ will be marked as a sink and not returned on the Pareto-optimal frontier. Therefore, the described BFS is sound.

If none of those number of states admit a solution to \problem, then none of them are sinks, so $(m_1, m_2, \dots, m_n)$ will be reached, and if $(m_1, m_2, \dots, m_n)$ does admit a solution to \problem, it will correctly be returned as a solution on the Pareto-optimal frontier. Thus, the described BFS is complete.
\end{IEEEproof}

\subsection{DFA Encoding Sizes for DISS} \label{appendix:encode_sizes}
When sampling a concept identifying a set of labeled examples, DISS algorithms are exponentially more likely to sample concepts with smaller \textit{size complexity}, i.e., number of bits required to represent the concept. We define the \textit{size} of a DFA decomposition $\mathcal{A} = (\mathcal{A}_1, \dots, \mathcal{A}_n)$ with number of states $m_1, \dots, m_n$ based on the size of the underlying DFAs as:
\begin{equation}
    \sum_{i=1}^n\text{size}(\mathcal{A}_i) - (n - 1)(2\ln(\Sigma) + 1) - 2(n - 1)\ln(m_1)
\end{equation}
where the size of each underlying DFA is given by $\text{size}(\mathcal{A}_i) = 3 + 2 \ln(m_i) + 2 \ln(\Sigma) + (|F_i| + 1) \ln(m_i) + z ( \ln(\Sigma) + 2 \ln(m_i) )$ where $F_i$ are the accepting states of $\mathcal{A}_i$ and $z$ is the number of non-stuttering transitions of $\mathcal{A}_i$.

\subsection{Extra Figures} \label{appendix:figs}

\begin{figure}[h]
    \centering
  \includegraphics[width=\linewidth]{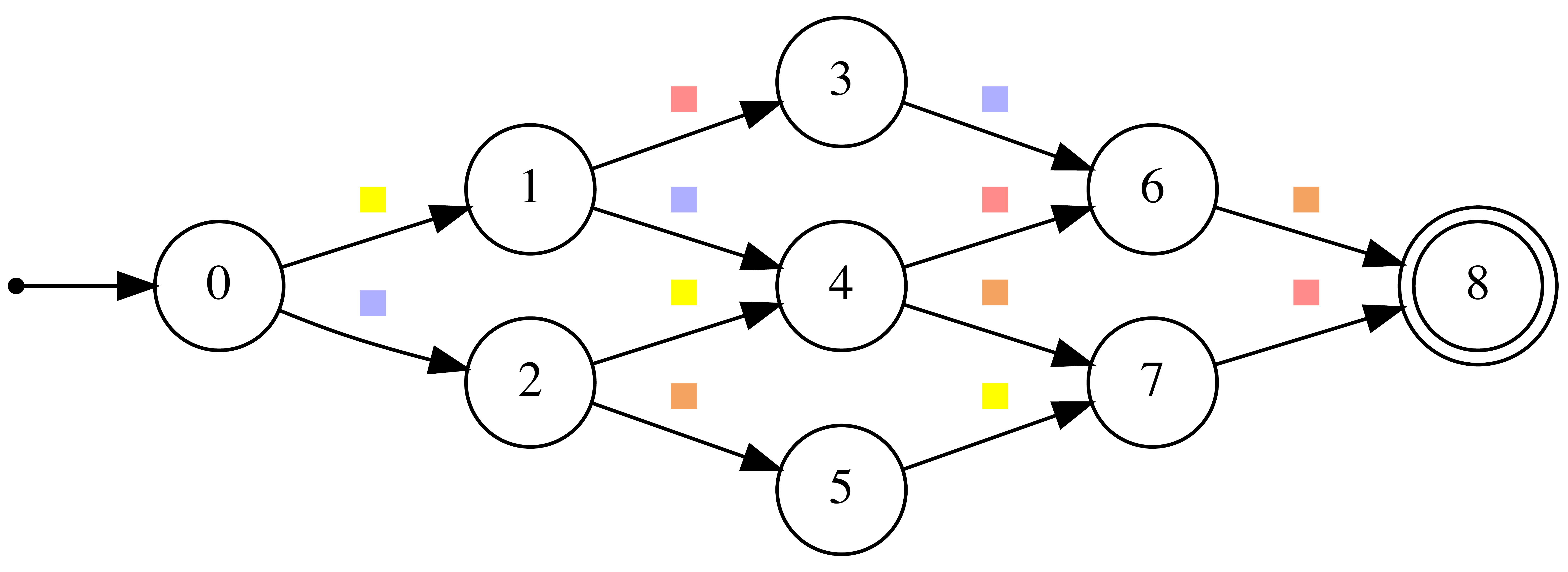}
  \caption{Monolithic DFA for the example presented in \Cref{subsec:toy_example}.}
  \label{fig:monolithic_dfa} 
\end{figure}



\begin{figure}[h]
    \begin{center}
        \scalebox{.325}{
        \centering \import{figures/}{mass_Monolithic_w_cbar_long.pgf} }
             \caption{How quickly DISS finds explanatory DFA decompositions compared to the enumeration baselines for a setting without prior partial knowledge. The temperature $\beta$ controls the degree of likelihood to which better explaining DFA decompositions are sampled. \label{fig:diss_exp:plot}
         }
    \end{center}
\end{figure}   

\begin{figure}[h]
    \begin{center}
        \scalebox{.325}{
        \centering \import{figures/}{mass_Incremental.pgf} }
             \caption{How quickly DISS finds explanatory DFA decompositions compared to the enumeration baselines for a setting with prior partial knowledge. The temperature $\beta$ controls the degree of likelihood to which better explaining DFA decompositions are sampled.\label{fig:diss_exp:inc_plot}
         }
    \end{center}
\end{figure}      

\begin{figure}[h]
    \centering
  \includegraphics[width=0.8\linewidth]{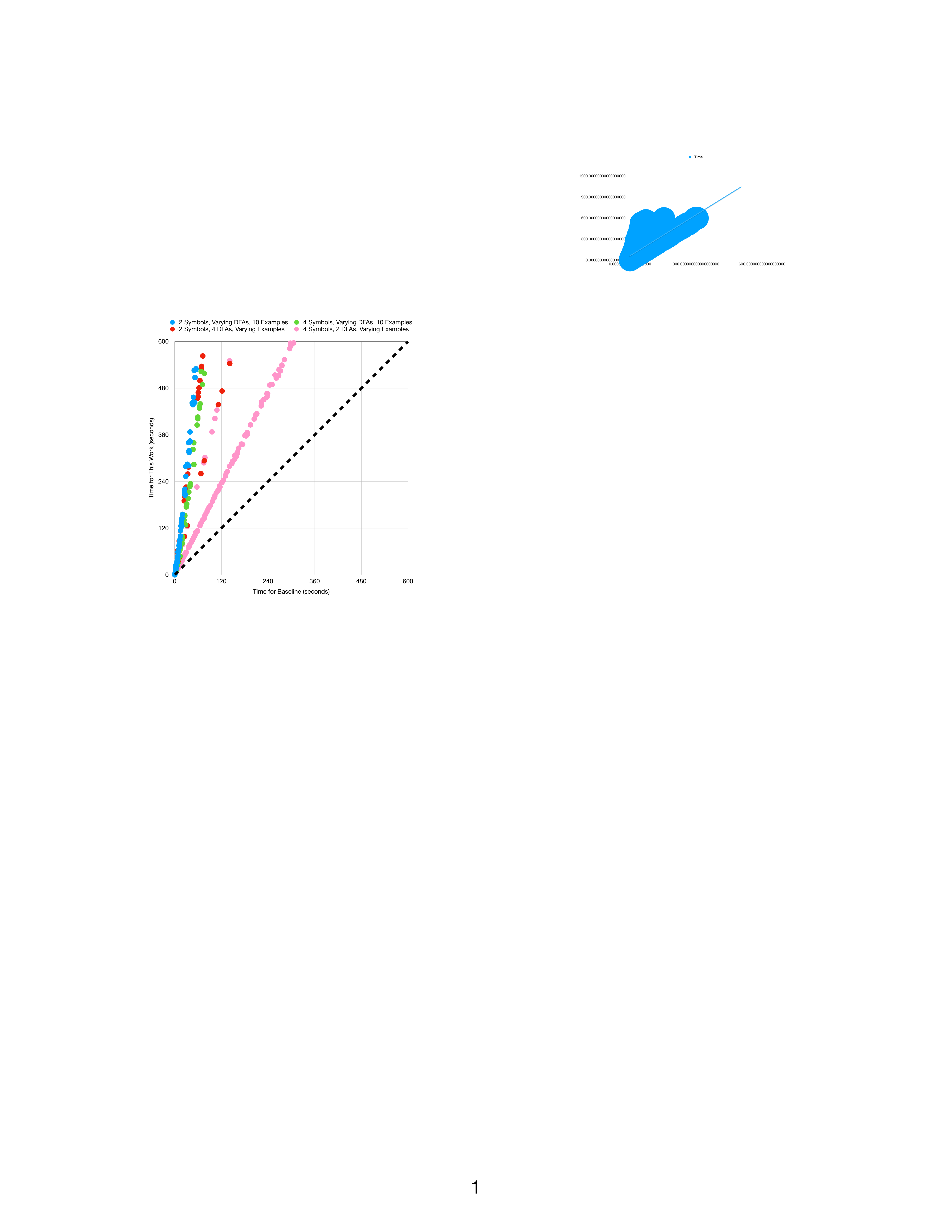}
  \caption{This plot compares the scalability of our algorithm to the monolithic DFA identification baseline. Each point in the plot represents a problem instance that is in the experiments presented in \Cref{subsec:dfa_eval}. The dashed black line is the $x = y$ line. As expected, our algorithm requires more time to solve same problem instances. However, we should note that the relationship between the solution times is not far away from the $x = y$ line.}
  \label{fig:all} 
\end{figure}

\end{document}